\documentclass[twocolumn,pra,aps,superscriptaddress,showpacs,flushbottom,floatfix]{revtex4-1}
\usepackage{xspace,amsmath,amsfonts,amsthm,amssymb,amsbsy,graphicx,color}
\newcommand{\ms}[1]{\mbox{\scriptsize #1}}

\begin{document}

\newtheorem{theo}{Theorem} \newtheorem{lemma}{Lemma}

%\title{Comparing coherent and continuous measurement-based feedback \\ for controlling a single qubit}
\title{Coherent vs.\ measurement-based feedback for controlling a single qubit}

\author{Ashkan Balouchi}
\affiliation{Hearne Institute for Theoretical Physics, Louisiana State University, Baton Rouge, LA 70803, USA} 
\author{Kurt Jacobs} 
\affiliation{U.S. Army Research Laboratory, Computational and Information Sciences Directorate, Adelphi, Maryland 20783, USA} 
\affiliation{Department of Physics, University of Massachusetts at Boston, Boston, MA 02125, USA}
\affiliation{Hearne Institute for Theoretical Physics, Louisiana State University, Baton Rouge, LA 70803, USA} 

\begin{abstract} 
We compare the performance of continuous coherent feedback, implemented using an ideal single-qubit controller, to that of continuous measurement-based feedback for the task of controlling the state of a single qubit. Here the basic dynamical resource is the ability to couple the system to a traveling-wave field (for example, a transmission line) via a system observable, and the fundamental limitation is the maximum rate that is available for this coupling. We focus on the question of the best achievable control given ideal controllers. To obtain a fair comparison we acknowledge that the amplification involved in measurement-based control allows the controller to use macroscopic fields to apply feedback forces to the system, so it is natural to allow these feedback forces to be much larger than the mesoscopic coupling to the transmission line that mediates both the measurement for measurement-based control and the coupling to the mesoscopic controller for coherent control. Interestingly our numerical results indicate that under the above platform for comparison, coherent feedback is able to exactly match the performance of measurement-based feedback given ideal controllers. We also discuss various properties of, and control mechanisms for, coherent feedback networks. 
\end{abstract} 


\maketitle 

\section{Introduction}
\vspace{-2mm}

Measurement-based feedback control is the process of making measurements on a quantum system, and using the results of the measurements to apply forces to the system to control it~\cite{Jacobs14, WM10, Kubanek09, Brandes10, Gillett10, Koch10, Sayrin11, Chen12, JZhang12, Vijay12, Brakhane12, Riste12, Naoki14, Balouchi14, Martin15, Cox16, Sudhir16, Jacobs10, Jacobs07c, Zhang16}. An alternative means of realizing feedback control is to have the system interact with a second ``auxiliary'' quantum system. The auxiliary quantum system can extract information from the primary system, via the interaction, and apply forces back onto the system, also via the interaction, thus implementing feedback control without the use of measurements. This paradigm for controlling quantum systems is referred to as \textit{coherent feedback control}~\cite{Zhang16, James08b, Gough09b, Nurdin09, Hamerly12, Hamerly13, Jacobs14b, Naoki14, Dong16}. Since many control processes that involve auxiliary systems are designed without reference to feedback processes, coherent feedback should be viewed more as a way of thinking about quantum control processed rather than as a distinct control technique. From a purely theoretical point of view coherent feedback subsumes control the uses measurements --- any measurement-based  process can be implemented in a coherent manner, technology permitting~\cite{Jacobs14b, Jacobs14}. 

The discovery by Nurdin, James, and Petersen~\cite{Nurdin09} that a coherent feedback process can outperform measurement-based feedback for linear quantum systems began a quest to understand the relationship between the two forms of control. While the difference discovered by Nurdin \textit{et al.\ \emph{was}} small, Hamerly and Mabuchi subsequently showed that coherent feedback could significantly outperform its measurement-based counterpart for cooling oscillators when the controller was restricted to linear interactions and controls~\cite{Hamerly12, Hamerly13, Jacobs15}. Following this it was shown in~\cite{Jacobs14b} that when the strength of the system/auxiliary interaction Hamiltonian is bounded, coherent feedback can significantly outperform measurement-based feedback even when the controller has access to arbitrary (non-linear) interactions and controls. In this case the difference between the two is due to a fundamental restriction on the paths in Hilbert space that measurement-based control can use. We note also, that in addition to the quantitive relationships between coherent and measurement-based control determined in the works mentioned above, Yamamoto~\cite{Naoki14} and Wiseman~\cite{Wiseman94} have also determined important qualitative differences between the two forms of control.

Here we consider a somewhat different, and arguably more experimentally relevant, constraint on control resources than that used in~\cite{Jacobs14b}. The forces that can be applied to a system, or more specifically the physical coupling between a system and an external controller, is an important and fundamental resource for control. A key question in quantum control is therefore how a constraint on this coupling affects the optimal control that can be achieved for the system. In~\cite{Jacobs14b}, and also in~\cite{Wang11}, the authors considered the best possible control that could be obtained when the norm of the coupling Hamiltonian with the system that is bounded. This particular choice of constraint on the coupling is most appropriate when the quantum system is finite dimensional and is coupled directly to another mesoscopic system that is also finite dimensional. In this case the coupling Hamiltonian is finite dimensional and the norm of this Hamiltonian characterizes well the maximum forces that the controller can apply to the system. 

A quite different, but also physically natural way to control a system is to couple it to a traveling-wave field, and such a field is effectively infinite dimensional. In this case the norm of coupling Hamiltonian may be unbounded, and no-longer characterizes the rate at which the controller can modify the state of the system. Instead the forces that can be applied by the controller, certainly in the limit in which the coupling to the field is broadband (Markovian) can be characterized solely by the system operator that couples the system to the field, and in particular by the norm of this operator. (Note that the overall size of any coupling between the system and field can always be absorbed into the system observable). It is a coupling to a traveling-wave field that is used to make a continuous measurement on a system, and it is therefore this type of coupling that is relevant for continuous-measurement based feedback control. A field-coupling can also be used to couple mesoscopic systems together and thus implement coherent feedback control. Here we compare the performance of measurement-based and coherent feedback control when the fundamental limitation is the magnitude of the coupling to the field. Specifically we will characterize this magnitude using a measure of the rate at which the field extracts information about the system (see below). More simply this can be thought of merely as the norm of the operator that appears in the master equation for the system when the field is traced out. We will refer to this norm as the \textit{strength} of the field-coupling.

In order to compare measurement-based feedback (MBF) to coherent feedback we take into account the following important difference between the two. To extract information from a mesoscopic system the system must interact with another system that is also mesoscopic, by which we mean it has a similar energy scale to the first. This is because in order for the second ``probe'' system to learn about the state of the first, the latter must change the state of the probe appreciably. That is, it must have an appreciable effect on probe. However, to reliably store and process the information obtained by the probe this information must, at least with present technology, be stored on circuits that have a much higher energy scale than the of typical mesoscopic quantum systems. The information in the probe must somehow be transferred to a much more macroscopic system, and this is the process of \textit{amplification}. Note that the process of amplification effectively allows a mesoscopic system (the probe) to have an appreciable effect on a macroscopic system. It is because this process is most readily performed in stages that motivates our previous assertion that the measurement must be initially realized by coupling the system to another mesoscopic system, rather than directly to a macroscopic one. As something of an aside, it is worth noting that the amplification involved in a measurement is the only part of the process of measurement that distinguishes measurement from any other quantum dynamical process~\cite{Cruikshank16}. 

The fact that the results of a measurement are amplified to a macroscopic level means that the mechanism by which MBF can apply feedback forces to a system is quite different than that available to coherent feedback control (CFC). Because coherent feedback must maintain coherence --- and thus quantum behavior --- during the feedback loop, all the control must be implemented using mesoscopic systems (at least given present technology), and thus the feedback forces must be applied using an interaction between mesoscopic systems. Therefore, if there is an experimental limitation on the strength that can be achieved between a mesoscopic system and a field, the feedback forces applied by CFC must be subject to this same limitation. The feedback forces that are applied by MBF, on the other hand, can be applied using fields with macroscopic amplitudes, and as a result are not subject to the same constraint as the strength of the coupling via which a mesoscopic system affects a field. The reason for this is that the interaction strength, or the force, by which the macroscopic field affects the system is proportional not only to the system operator that couples to the field, but also to the ``coherent-state'' amplitude of the field. Thus while the forces applied by a mesoscopic system to a field are ``weak'' those applied by the field to the system can be ``strong'' if the field has sufficient amplitude. 

Given the above discussion of the physical difference between MBF and CFC, we conclude that a fair comparison between the two is obtained, at least for the purposes of current experimental technology, by placing the same limit on the strength with which the measurement component of MBF interacts with a given system as that in which a CFC controller interacts with the system, but allowing the feedback forces applied by MBF to be as large as desired. That is the basis we will use for our comparison here. 

In the next section we define the task, or ``control problem'', for which we will compare MBF and CFC, and define what we mean by ``perfect'' or ``ideal'' controllers''. In Section~\ref{bounds} we define precisely how we quantify the strength of a Markovian coupling between a system and a field, and thus how we quantify the constraint in our control problem for both classical and quantum controllers. We also explain why we restrict the coupling between the system-to-be-controlled and the fields to those in which the coupling operator is Hermitian. In Section~\ref{Hpic} we briefly review the Heisenberg-picture quantum noise equations that we use to describe the coupling between the systems and the fields, and introduce some useful notation. In Section~\ref{secmbf} we describe the control of a single qubit with a continuous measurement and discuss briefly what is known about the performance of this kind of control. In particular, we review the optimal performance of this control method in the limit in the which the feedback force is infinite, which has been established in previous work, and present numerical results on the optimal performance when the feedback force is finite. In Section~\ref{cfc} we introduce the CFC configurations that we consider, which cover all configurations in which both the system and controller have two interactions with a field, and discuss some of there properties. In Section~\ref{numcfc} we use numerical optimization to explore the performance of these configurations when the controller is a single qubit, and we compare this performance to that of continuous measurement-based control described in Section~\ref{secmbf}. Finally, we summarize the results and some open questions in Section~\ref{conc}, and the Appendix presents further details of the method we use for the numerical optimization.

\vspace{-3mm}
\section{The task}
\vspace{-2mm}

Here we consider the control of a single qubit, which provides not only a system that is experimentally relevant, but also one that is relatively simple dynamically and thus a good platform for evaluating the relative performance of various control systems. As the  task for our controllers we choose that of maintaining the qubit in its ground state in the presence of thermal noise. This task if thus one of steady-state control. As our measure of performance (strictly, lack of performance) we choose the steady-state probability $P$ that the qubit is in its excited state. Denoting the ground and excited states of the qubit by $|0\rangle$ and $|1\rangle$, respectively, the master equation for the qubit in the absence of any control is 
\begin{equation}
 \dot{\rho} =  -i\left[\frac{H}{\hbar},\rho\right] - \frac{\gamma}{2} \left[ (n_T + 1) \mathcal{K}(\sigma) + n_T \mathcal{K}(\sigma^\dagger) \right] \rho 
 \label{therm}
\end{equation}
in which the Hamiltonian is $H = \hbar \Omega |1\rangle \langle 1|$,  $\gamma$ is the thermal relaxation rate of the qubit, and the superoperator $\mathcal{K}$ is defined by 
\begin{equation}
  \mathcal{K}(c) \rho = c^\dagger c \rho + \rho c^\dagger c - 2 c \rho c^\dagger 
\end{equation}
for an arbitrary operator $c$. The temperature of the bath is characterized by the parameter $n_T$ which is given by  
\begin{equation}
   n_T =\frac{1}{1-\exp(-\hbar\Omega/[k_{\ms{B}}T])}, 
\end{equation}
where $k_{\ms{B}}$ is Boltzmann's constant and $T$ is the temperature. In the absence of any control the steady-state (thermal) population of the excited state is 
\begin{equation}
   P_{\ms{therm}} = \frac{n_T}{1 + 2n_T} \approx n_T  , \;\;\;  n_T \ll 1 . 
\end{equation}  

\subsection*{Assumption of perfect controllers}

In our analysis here we assume that the controllers are perfect, since we interested in the best control that can be achieved by both methods when the only constraint is the speed of the interaction with the system. This means specifically that, for measurement-based control we assume that the measurement is perfectly efficient, meaning that there is no additional noise on the measurement result over the noise which is purely a result of the uncertainty inherent in the quantum state of the system. In addition we assume that the feedback forces applied by the classical control system have no errors, and the processing of the measurement results is effectively instantaneous. 

For coherent feedback control, since the controller is an auxiliary quantum system, the assumption of a perfect controller means that the auxiliary does not feel any thermal noise or decoherence beyond that which we choose to maximize its ability to effect control. That is, we are able to completely isolate it from the environment. While the interaction of the auxiliary with the traveling-wave fields that it uses to talk the system is subject to the same bound as the system, for consistency, the internal dynamics of the controller is unrestricted, which is equivalent to the instantaneous processing allowed by the classical controller. Finally, the traveling-wave fields that connect the system and the auxiliary are assumed to have no loss, which is equivalent to the assumption that the measurements made by the classical controller are perfectly efficient. 

\section{The measurement rate as the constrained resource}
\label{bounds}

A continuous measurement of an observable $A$ of a quantum system is obtained by coupling the system to a probe system via $A$ and coupling the probe to a traveling-wave field. The reason that we use this two-stage process for coupling a system to a field, and thus for making a continuous measurement, as opposed to coupling the system directly to the field is the following. To obtain a simple Markovian process whereby the field continuously carries information away from the probe the frequency of the photons emitted by the probe must be large compared to the rate at which the probe emits photons into the field. The emitted field then contains a signal whose bandwidth is small compare to the frequency of the photons, and it is within this bandwidth that the field must cary the signal containing the information about the measured observable of the system. We can achieve this using by a probe to mediate the interaction because we can choose the frequency of the probe  to be significantly larger than the timescale of the evolution of the system. Explicit treatments of this continuous measurement process can be found in~\cite{JacobsSteck06, Jacobs14, WM10}. The result of the coupling to the probe and the subsequent coupling of the probe to the field is a master equation that describes the evolution of the system given the continuous stream of measurement results obtained by detecting the field. If we denote this stream by $r(t)$ the master equation for the system density matrix, $\rho$, is 
\begin{align} 
    d\rho = & -i\left[\frac{H_{\ms{s}}}{\hbar},\rho \right] dt - k [\tilde{A},[\tilde{A},\rho]] dt \nonumber \\ 
                & + \sqrt{2 k} ( \tilde{A}\rho + \rho \tilde{A} - 2\mbox{Tr}[\rho \tilde{A}] \rho) dW .            
    \label{sme}       
\end{align} 
Here we have written $A = \sqrt{k} \tilde{A}$, so that if $\tilde{A}$ is a dimensionless operator $k$ is a rate-constant. We will find this definition useful below. In the above master equation, $H_{\ms{s}}$ is the Hamiltonian of the system and $dW$ is the stochastic increment of Weiner noise. This noise is the random fluctuations in the stream of measurement  results, and is given by $dW(t) = (r(t) - \mbox{Tr}[A]\rho) dt$. The stronger the coupling between the system and field, mediated by the probe, the larger are the eigenvalues of $A$, and thus the larger is $k$ if $\tilde{A}$ is fixed.  

The rate at which information can be extracted from the system is the resource in which we are interested, a resource that is ultimately determined by the strength of the (indirect) coupling between the system and a traveling-wave field. In particular we wish to know how well a control system can perform when the rate at which information can be extracted is the limiting factor, meaning that all other resources are unlimited. This rate is easily characterized using the difference between the two eigenvalues of $A$ --- the larger this difference the faster the master equation above projects the system into one of the eigenstates of $A$ (in the absence of any Hamiltonian dynamics). We also note that the difference between the eigenvalues of a two-dimensional Hermitian operator is a perfectly good norm for such operators, and so we will denote this difference by $|A|$.  Further, if $A$ were to appear in the Hamiltonian of a qubit, then $|A|$ measures the maximum speed at which $A$ can generate evolution in Hilbert space~\cite{Anandan90, Margolus98, Carlini06, Carlini07, Wang13, Wang15}. This quantity that provides a good measure of the strength of the feedback force for MBF. 

We must now place a constraint on the coupling between the system and a field, a coupling that will always be implemented using the two-stage procedure describe above. Let us say that the system is coupled to the field via an interaction with the probe that is proportional to an Hermitian system operator $X$, so that $X$ is the Lindblad operator that appears in the master equation as the result of this coupling. As such, the dimensions of $X$ are that of the square root of a rate. We apply the constraint by writing  
\begin{align} 
    X = \sqrt{k} \tilde{X}                   
\end{align} 
where $k$ is a rate constant and $\tilde{X}$ is dimensionless. This allows us to define the constraint on our resource by fixing the value of $k$ and imposing the bound 
\begin{align} 
    |\tilde{X}|  \leq  2 .                   
\end{align} 
The value of $k$ thus sets the bound on any field-coupling with the system that the controller can exploit. 

\subsection*{Restriction to Hermitian coupling operators} 

We restrict our control protocols to use only couplings between the system and a field in which the Lindblad operator in the resulting master equation is Hermitian. (In Eq.(\ref{sme}) above the Lindblad operator is $A$).  One reason for this is that we are interested in comparing coherent control protocols with measurement-based protocols that use a measurement of an observable. The reason that we focus on measurements of an observable has to do with the fact that a crucial part of a control processes is that of extracting entropy from a system. Processes described by Lindblad operators that are not Hermitian describe dissipative processes. Such processes are able to extract entropy themselves without requiring any feedback or other control mechanism. For example, if we have a system driven by thermal noise and we are allowed to couple that system to a zero-temperature bath, which is a purely dissipative coupling, then we obtain a significant amount of control for free, without having to construct a control mechanism that utilizes information obtained about the system. It is for this reason that we are interested in comparing coherent protocols that are restricted to non-dissipative interactions with measurement-based protocols that are similarly restricted.  

\section{The Heisenberg picture and output fields}
\label{Hpic}

The stochastic master equation (SME) given in Eq.(\ref{sme}) is useful for describing feedback control via a continuous measurement, but the Heisenberg picture is most convenient for describing the situation in which the quantum field that caries information away from the system is coupled to another system in order to implement coherent feedback control. In the Markovian regime of broad-band coupling to the field it is possible to write Heisenberg equations of motion for the system that are driven by the field, and then to write the field after it has interacted with the system in terms of the original field and a contribution from the system~\cite{Collett84, Gardiner85}. The Heisenberg equations of motion for an arbitrary system operator $S$ that correspond to the master equation above are~\cite{Jacobs14, Gardiner10} 
\begin{align} 
    dS = & \; i\left[\frac{H_{\ms{s}}}{\hbar},\rho \right] dt  - k \mathcal{K}(\tilde{A})S dt \nonumber \\  
         & +  \sqrt{2 k} ( [S,\tilde{A}^\dagger] da_{\ms{in}} - [S,\tilde{A}] da_{\ms{in}}^\dagger ) 
\end{align} 
where we have now written $A = \sqrt{k} \tilde{A}$ and $da_{\ms{in}}$ is an increment of the quantum operator that describes the input field. This field operator has properties that are analogous to Weiner noise and in particular $[da_{\ms{in}},da_{\ms{in}}^\dagger] = dt$ for a vacuum input. Note that if $c$ is Hermitian then $\mathcal{K}(c)S = [c,[c,S]]$. Since the decoherence term given by the super-operator $\mathcal{K}$ must always appear with the terms containing the input field it is convenient to introduce a more compact notation for these terms. Noting that we can multiply the operator $A$ in the Heisenberg equation of motion by a phase factor $e^{-i\theta}$ without changing the superoperator $\mathcal{K}$, and that this is equivalent to applying the phase shift $e^{i\theta}$ to the input field $a_{\ms{in}}$, we define a new superoperator $\mathcal{Q}$ by
%\begin{widetext}
\begin{eqnarray} 
    d\mathcal{Q}[A, \theta, a_{\ms{in}}] S & \equiv & - \mathcal{K}(A)S dt  \nonumber \\  
    & & +  \sqrt{2} \left[ S, A^\dagger  e^{i\theta} da_{\ms{in}} - A e^{-i\theta} da_{\ms{in}}^\dagger \right] \!\! . 
\end{eqnarray} 
%\end{widetext}
The field that is output from the system --- that is, the traveling-wave field \textit{after} it has interacted with the system --- can be written as 
\begin{align} 
     da_{\ms{out}} = da_{\ms{in}} - \sqrt{2} A dt = da_{\ms{in}} - \sqrt{2k} \tilde{A} dt. 
\end{align} 
One of the beauties of this ``input-output'' formalism for describing systems interacting with fields is that we can easily describe the situation in which a field that has interacted with one system subsequently interacts with another system merely by setting the field operator that drives the second system equal to the output field operator from the first system. We will do this explicitly in what follows. 

\begin{figure}[t] 
\leavevmode\includegraphics[width=1\hsize]{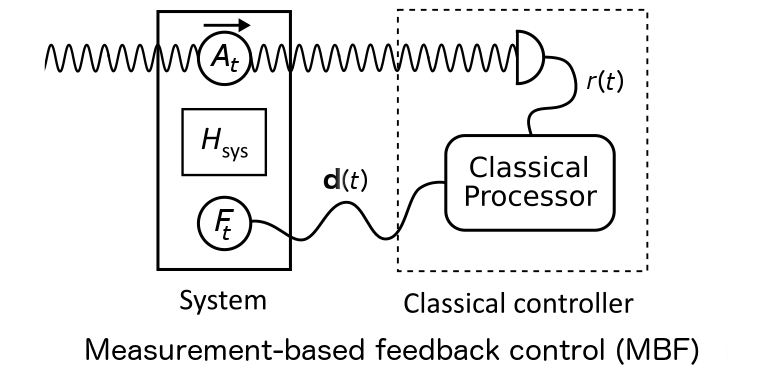} 
\caption{Here we depict a measurement-based feedback (MBF) control system. A traveling-wave field, represented by the wavy lines, interacts with the system to be controlled via an observable, $A$, that may be time dependent. The arrow denotes the direction in which the field carries away the output from the system. The field is measured continuously by a detector that produces the classical measurement signal $r(t)$. This classical signal is processed by a computing device of some sort and the result is another continuous-time signal, $\mathbf{d}(t)$, that may be vector valued. The latter is used to apply forces to the system by adding a term $H_{\ms{mbf}}(t) = \hbar F[\mathbf{d}(t)]$ to the system Hamiltonian, in which $F$ is a Hermitian operator that depends on $\mathbf{d}$. } 
\label{fig1} 
\end{figure} 

\section{Feedback control via continuous measurement}
\label{secmbf}

In Fig.~\ref{fig1} we depict a feedback control system that uses continuous measurement. Here a field traveling to the right, which might physically be an optical beam or superconducting transmission line, interacts with the system via a system observable $A(t)$ and is subsequently measured. The continuous classical measurement signal, denoted by $r(t)$, is processed by a classical computing device to produce a vector-valued signal $\mathbf{d}(t)$. This signal is then used to control the system by applying control fields to it. The Hamiltonian of the system becomes 
\begin{equation}
   H = H_{\ms{s}} + \hbar F[\mathbf{d}(t)] 
\end{equation}
in which $H_{\ms{s}}$ is the time-independent Hamiltonian of the system in the absence of the control, and the Hermitian operator $F[\mathbf{d}(t)]$ describes the control forces that are applied at time $t$. 

\begin{figure}[t] 
\leavevmode\includegraphics[width=1\hsize]{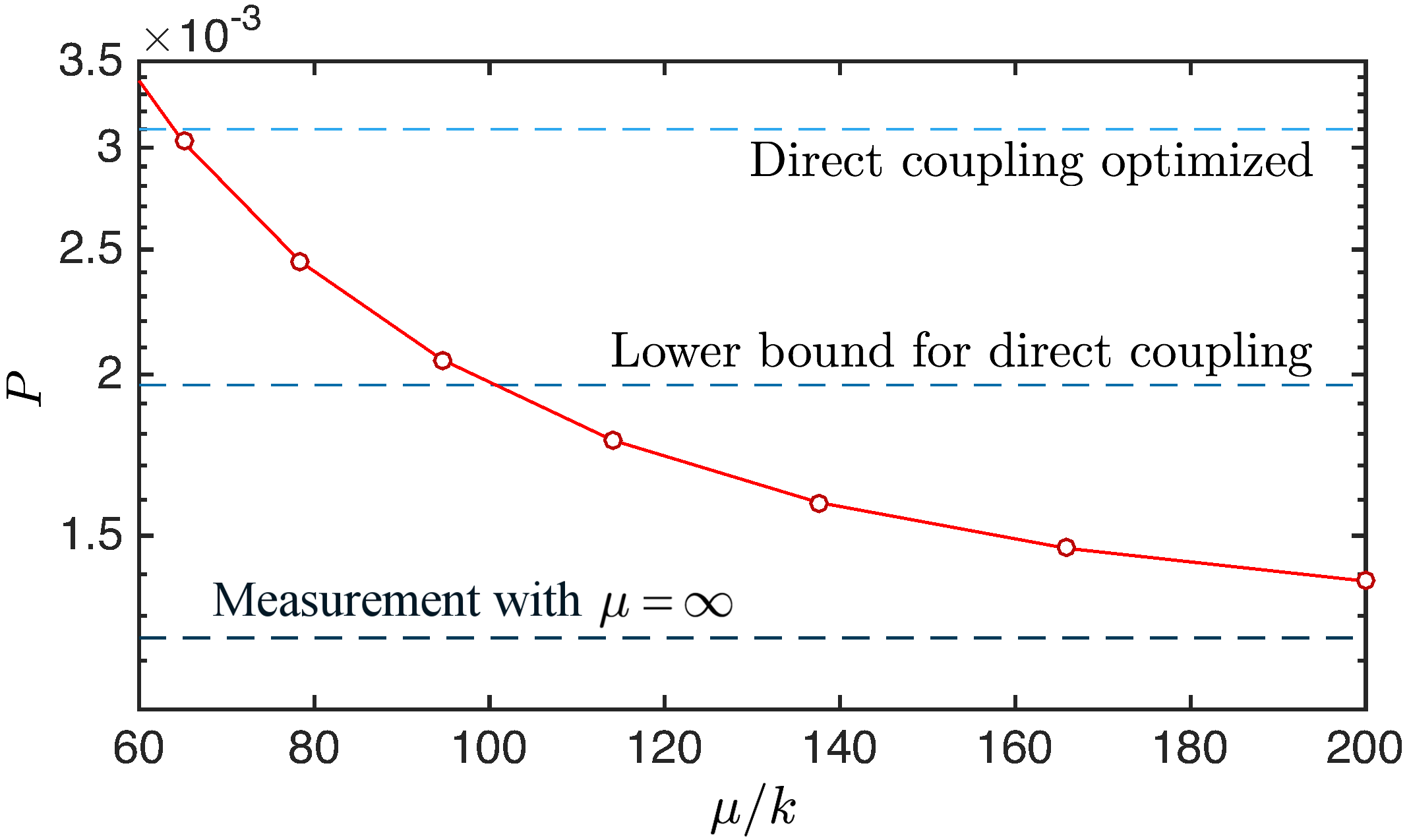} 
\caption{Here the solid line shows the performance of a measurement-based feedback protocol that is optimal for preserving the ground state in the presence of thermal noise when $\mu \gg k$, where $\mu$ is the control speed and $k$ is the measurement strength. Here $P$ is the quantity to be minimized, being the excited-state probability in the steady-state. In the absence of control $P = 1/3$. The thermal relaxation rate is $\gamma = 0.01 k$ and the parameter that defines the temperature of the thermal bath is $n_T = 1$ (see text). The dashed lines provide comparisons with various quantities of interest (see text).} 
\label{fig2} 
\end{figure} 

The dynamics of the system under the MBF feedback scenario shown in Fig.~\ref{fig1} is described by the following stochastic master equation (SME)~\cite{Jacobs14, WM10, JacobsSteck06}
\begin{align} 
    d\rho = & -i\left[\frac{H_{\ms{s}}}{\hbar} + F[\mathbf{d}(t)],\rho \right] dt - k [\tilde{A},[\tilde{A},\rho]] dt \nonumber \\ 
                & + \sqrt{2 k} ( \tilde{A}\rho + \rho \tilde{A} - 2\mbox{Tr}[\rho \tilde{A}] \rho) dW + \mathcal{L}_{\ms{env}} \rho .               
                \label{mbf}    
\end{align} 
Here $\rho$ is the density matrix for the system, $dW$ is an increment of white Gaussian noise, referred to as Wiener noise~\cite{Jacobs10c}, and $\mathcal{L}_{\ms{env}}$ is a linear superoperator that describes the effect of additional noise to which the system is subjected. The control Hamiltonian $H_{\ms{mbf}} = \hbar F[\mathbf{d}(t)]$ can in general be chosen to be any function of the measurement record $r(t')$ up until the current time $t$. We will quantify the strength (or speed) of control using $|F|/2$, and denote the limit to this strength by $\mu$, meaning that   
\begin{align} 
 \frac{|F(t)|}{2} \leq \mu . 
\end{align} 

We now review the level of control that can be obtained with continuous-time measurement-based feedback, as  described by Eq.(\ref{mbf}). The question of the optimal MBF control protocol for a single qubit is remarkably complex, and is not known in general. However an optimal protocol is known in the limit of strong feedback ($\mu/k \rightarrow \infty$)~\cite{Shabani08, Jacobs08b, Jacobs07c}. This protocol involves using the measurement record to continually calculate the density matrix $\rho(t)$, and continually modifying the feedback Hamiltonian $F(t)$ and measured observable $A(t)$ so that i) in each time-step $dt$ the total Hamiltonian rotates the Bloch vector towards the target state (in our case the ground state $|0\rangle$) as fast as possible, and ii) the eigenbasis of the measured observable remains unbiased with respect to that of the density matrix. This second condition can also be stated by saying that the Bloch vectors of the eigenstates of $A(t)$ remain orthogonal to those of the eigenstates of the density matrix. 

Under the above MBF protocol, in the limit in which $\mu \rightarrow \infty$ and in the absence of any noise, the purity, $\mathcal{P}$, of the system density matrix evolves as $d\mathcal{P}/dt = -8k\mathcal{P}$, and the density matrix remains diagonal in the $Z$-basis. Adding the thermal noise described by the master equation in Eq.(\ref{therm}) it is simple to calculate the resulting steady-state value of the excited-state probability, which is 
\begin{align} 
 P_{\ms{MBF}}^{(\infty)} = \left( \frac{\gamma n_T}{8k}  \right) .   % \;\;\;\; k \gg \gamma n_T   
 \label{infmu}
\end{align} 

The results in~\cite{Balouchi14} indicate that the above feedback protocol remains optimal so long as $\mu \gtrsim 50 k$, but is it not possible to obtain an analytic solution for the performance for general $\mu$. We therefore simulate the protocol numerically, for which we choose the values $0.01 k$ for the thermal relaxation rate $\gamma$ and $1$ for the temperature parameter $n_T$. 
Note that the control problem is defined by four parameters, and only the three parameters $\gamma$, $n_T$, and the feedback strength $\mu$ if we scale everything by the measurement rate $k$ (that is, measure all rates in units of $k$). We plot the performance of the protocol in Fig.~\ref{fig2} as a function of $\mu/k$. For these simulations we used the recent numerical method devised by Rouchon and collaborators~\cite{Rouchon15, Amini11}, which is a tremendous advance on previous methods for simulating the SME, both in terms of stability and accuracy. 

\section{Coherent feedback control}
\label{cfc}

\begin{figure}[t] 
\leavevmode\includegraphics[width=0.9\hsize]{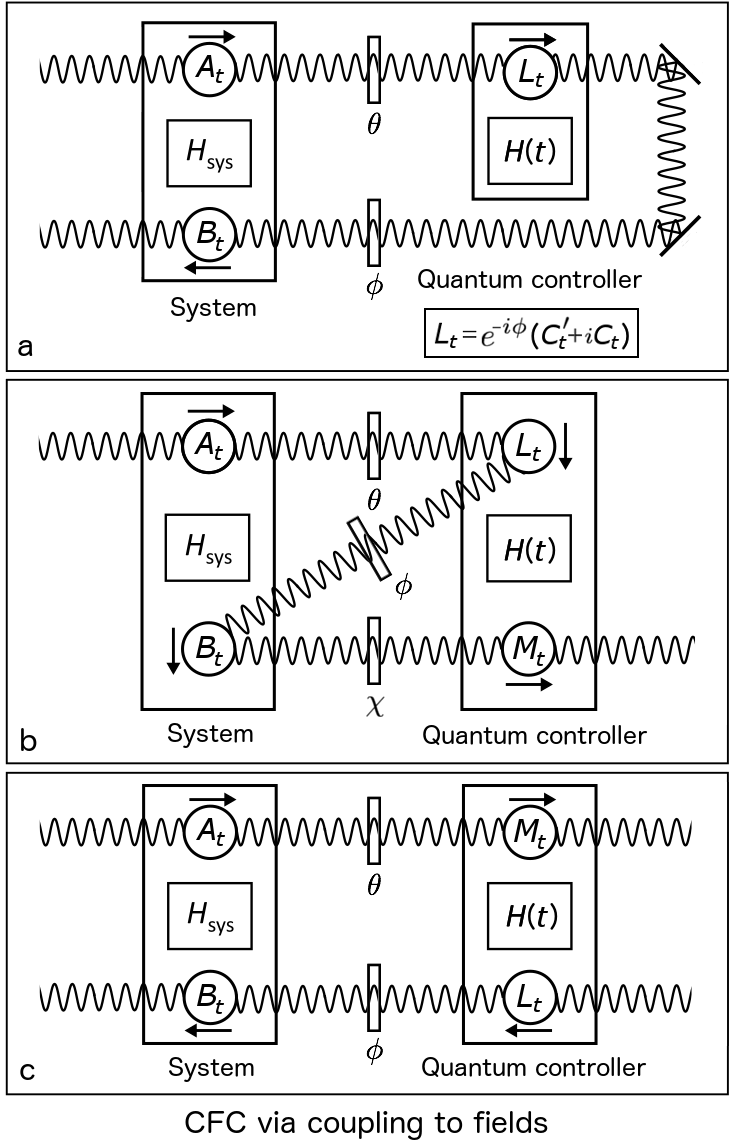} 
\caption{Here we show three ways in which coherent control can be implemented when the system and auxiliary each have two interactions with traveling-wave field(s). While we have separated configurations (a) and (b) for clarity, note that (a) is merely (b) with $M=0$. The field(s) are represented by the wavy-lines and interact with the systems via the operators in the circles. The arrows indicate the direction in which the field(s) carry information away from the systems and $\theta$ and $\phi$ denote phase shifts applied to the field(s).  The operators $A$ and $B$ are Hermitian while $L$ and $M$ may be non-Hermitian, with the subscripts indicating possible time-dependence. It is convenient to write an arbitrary $L$ in terms of Hermitian operators $C$ and $C'$ and the phase shift $\phi$.} 
\label{fig3} 
\end{figure}

To compare CFC protocols with the MBF protocol depicted in Fig.~\ref{fig1} the CFC protocols must interact with the system via traveling-wave fields. Three possible configurations that we can use to implement field-mediated CFC are shown in Fig.~\ref{fig3}.  In configurations (a) and (b) the system and controller are connect by a single traveling-wave field. In (a) this field interacts with the system via $A(t)$, goes on to interact with the controller (a second quantum system), and then returns to interact with the system via an operator $B(t)$ before passing out to be discarded. The only difference between (a) and (b) is that after the field has provided feedback to the system via operator $B$ it is then allowed to interact with the controller once more before being discarded. Phase shifters are also included that can apply shifts $\theta$ and $\phi$ to the field. Configuration (c) is different from the others in that two distinct fields are used. The field that carries the control signal from the controller to the system is not the field that carries information the other way. In all three configurations the interaction operators that mediate interactions between the system and controller must be bounded as $A$ in the MBF protocol. Further, we should not allow the system to be damped arbitrarily by the field; since the field is at zero temperature such damping would provide an entropy dumping mechanism for free that we did not allow in the MBF protocol. This condition can be imposed merely by demanding that $A$ and $B$ are Hermitian as we did for the MBF protocol. We can allow the controller as much damping as we want, however, and so the operators $L$ and $M$ can be non-Hermitian. We will decompose $L$ into Hermitian operators by writing $L = e^{-i\phi}(C' + iC)$. With this decomposition we will find that it is only $C$ that appears in the resulting field-mediated interaction with the system. Thus while $C$ must be subject to the same bound as $A$ in the MBF protocol, $C'$ can be left unbounded since it does not play any role in mediating a mesoscopic interaction. For the same reason $M$ is unbounded. We are also free to include additional arbitrary damping channels for the controller if we wish. 

For configuration (a) the Heisenberg equations of motion for an arbitrary system operator, $S$, and an arbitrary auxiliary operator, $X$, are given by~\cite{Jacobs14} 
\begin{align} 
    d S = & \, i \! \left[ \frac{H_{\ms{s}}}{\hbar} + 4 k \tilde{B}  \tilde{C}, S \right] dt  + d\mathcal{Q}[\sqrt{k}\tilde{D}, 0, a_{\ms{in}}] S , \label{S1a}  \\  
%      &- k \mathcal{K}(\tilde{D}) S dt  + \sqrt{2k} ( [S, \tilde{D}^\dagger ] da_{\ms{in}} - \mbox{H.c.} ) \label{S1a}  \\  
%    d X =  & \, i \! \left[ \frac{H_{\ms{a}}}{\hbar} + 2 k \tilde{A} i \left(  e^{-i\theta} L  - e^{i\theta} L^\dagger \right), X \right] dt  - k \mathcal{K}(\tilde{L}) X dt + \sqrt{2k}(  [X, e^{i\theta}\tilde{L}^\dagger] da_{\ms{in}}  - \mbox{H.c.} ) . 
   d X =  & \, i \! \left[ \frac{H_{\ms{a}}}{\hbar} + 4 k \tilde{A}\, \mbox{Im}[e^{-i\theta}\tilde{L}] , X \right] \! dt  \nonumber \\ 
   & + d\mathcal{Q}[\sqrt{k}\tilde{L}, \theta, a_{\ms{in}}] X ,     \label{X1a}
 %  & - k \mathcal{K}(\tilde{L}) X dt + \sqrt{2k}(  [X, e^{i\theta}\tilde{L}^\dagger] da_{\ms{in}}  - \mbox{H.c.} ) .  \label{X1a}
\end{align} 
in which  
\begin{align} 
    \tilde{D} \equiv & \, \tilde{A} +  e^{-i(\theta+\phi)}\tilde{B} , \\ 
    \tilde{L}  \equiv &\, e^{-i\phi}(\tilde{C}' + i\tilde{C}) ,  \\  
    \mbox{Im}[e^{-i\theta}\tilde{L}] = &  \sin(\theta \! +\! \phi) \tilde{C}' - \cos(\theta \! +\! \phi) \tilde{C} . 
\end{align} 
For configuration (b) the equations of motion for the system are identical to those for (a), while the equation of motion for the auxiliary becomes 
\begin{align}  
    d X =  & \, i \! \left[ \frac{H_{\ms{a}}}{\hbar} + 2 k i \left( A e^{-i\theta} \tilde{V} + B e^{-i\chi} \tilde{M}  -  \mbox{H.c} \right), X \right] dt  \nonumber \\ 
    & + d\mathcal{Q}[\sqrt{k}\tilde{V}, \theta, a_{\ms{in}}] X  \label{Xb}
  %  & - k \mathcal{K}(\tilde{J}) X dt + \sqrt{2k}(  [X, e^{i\theta}\tilde{J}^\dagger] da_{\ms{in}}  - \mbox{H.c.} ) . 
\end{align} 
in which  
\begin{align} 
   \tilde{V} \equiv \tilde{L} + e^{-i(\phi+\chi)} \tilde{M} . 
\end{align} 
In Eq.(\ref{Xb}) the term ``H.c.''  represents the Hermitian conjugate of the expression that appears before it within the parentheses. Thus $(Y + \mbox{H.c.}) \equiv (Y + Y^\dagger)$. 
The equations of motion for (c) are 
\begin{align} 
    d S = & \, i \! \left[\frac{H_{\ms{s}}}{\hbar} + 4 k \tilde{B} \tilde{C}, S \right] dt  + d\mathcal{Q}[\sqrt{k}\tilde{A}, 0, a_{\ms{in}}]  S \nonumber \\ 
        & + d\mathcal{Q}[\sqrt{k}\tilde{B}, \phi, b_{\ms{in}}]  S ,  \\ 
    % - k[\mathcal{K}(\tilde{A}) + \mathcal{K}(\tilde{B})]S dt \nonumber \\ 
   %  & + \sqrt{2k}[S,\tilde{A}](da_{\ms{in}} -  da_{\ms{in}}^\dagger) \nonumber \\ 
   %  & + \sqrt{2k}([S,e^{i\phi}\tilde{B}] db_{\ms{in}} - \mbox{H.c.}) ,  \\ 
d X =  & \, i \! \left[\frac{H_{\ms{a}}}{\hbar}+ 2 k \tilde{A} \left( e^{i\theta} \tilde{M}^\dagger  - e^{-i\theta} \tilde{M} \right),X \right] dt  \nonumber \\ 
  & + d\mathcal{Q}[\sqrt{k}\tilde{L}, 0, b_{\ms{in}}] X  + d\mathcal{Q}[\sqrt{k}\tilde{M}, \theta, a_{\ms{in}}] X. 
 % & - k[\mathcal{K}(\tilde{M}) + \mathcal{K}(\tilde{L})]X dt \nonumber \\ 
  % & + \sqrt{2k}([X,\tilde{L}^\dagger db_{\ms{in}} + e^{i\theta}\tilde{M}^\dagger da_{\ms{in}} ]-   \mbox{H.c.}) , 
\end{align} 
By examining the equations of motion for (a) we see that the field mediates an effective coupling between the two systems. Specifically, in Eq.(\ref{S1a}) a Hamiltonian term proportional to the product of $\tilde{B}$ and $\tilde{C}$ appears. This effective Hamiltonian is generated by the field that flows from the auxiliary operator $L$ to the system operator $B$, and we note that it does not appear in the equation of motion for the auxiliary. Instead the auxiliary sees an effective Hamiltonian proportional to $A$ and $L$ that comes from the field that flows from the system to the auxiliary. The equations of motion are thus asymmetric in a way that a purely Hamiltonian coupling never is. 

In configuration (a) it is the fact that the \textit{same} field interacts with the system via $A$ and $B$ that causes the system to see an effective interaction with the field given by the Lindblad operator $D$.  Note that in (c) where separate fields interact with $A$ and $B$ the decoherence due to the fields must instead be described by two separate superoperators $\mathcal{K}(\tilde{A})$ and $\mathcal{K}(\tilde{B})$. The fact that the same field interacts with the system twice allows the noise introduced at the second input to cancel that at the first input (assuming no significant time-delay between the two points of input). This cancellation can be achieved, for example, by choosing the phase shifts so that $e^{i(\theta + \phi)} = -1$ and $A = B$, with the result that $D = 0$. This possibility makes configurations (a) and (b) very different from (c). 

\subsection{Properties of the CFC configurations} 

\begin{figure}[t] 
\leavevmode\includegraphics[width=1\hsize]{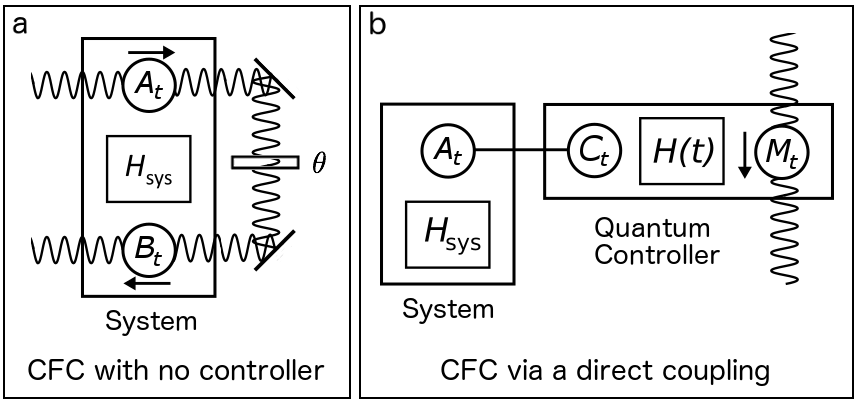} 
\caption{Here we show two configurations for implementing coherent feedback control. (a) This scenario is the special case of that in Fig.~\ref{fig3}(a) in which the auxiliary system has been removed. The processing of the output is merely a phase shift $\theta$ that is applied before the field is fed back to the system. (b) In this scenario, which is subsumed by that in Fig.~\ref{fig3}(a), the system is coupled directly to the auxiliary via the interaction Hamiltonian $H_{\ms{int}} = 4\hbar k \tilde{A}\tilde{C}$, which is indicated diagrammatically by the horizontal line. The auxiliary is also coupled to an output field to which it can discard entropy. These two configurations elucidate two control mechanisms that are available to the configurations in Fig.~\ref{fig3} (a) and (b).}  
\label{fig4} 
\end{figure}

We now examine some of the properties of the three CFC configurations in Fig.~\ref{fig3} so as to gain some insight into the possible control mechanisms. First, as we noted above, configuration (a) and (b) have the potential to cancel the noise that is fed into the system. If we take (a) and set $e^{i(\theta + \phi)} = -1$ and $A = B$ (or equivalently $e^{i(\theta + \phi)} = 1$ and $A = -B$) the noise terms arising from the two interactions with the field cancel, and the equations of motion become 
\begin{align} 
    \dot{S} = & \, i[H_{\ms{s}}/\hbar + 4 k \tilde{B}  \tilde{C}, S] ,   \\  
    d X =  & \, i[H_{\ms{a}}/\hbar  + 4 k \tilde{B} \tilde{C}, X ] dt  + d \mathcal{Q} [ \sqrt{k} \tilde{L}, \theta, a_{\ms{in}}  ]X . 
    % - k \mathcal{K}(L) X dt   \nonumber \\ 
            %  & + \sqrt{2k}( [X, e^{i\theta}\tilde{L}^\dagger] da_{\ms{in}}  - \mbox{H.c.} ) .  
     \label{halfH}
\end{align} 
Interestingly when we cancel the input noise for the system both the system and the controller see the same Hamiltonian interaction, which is given by $H_{\ms{eff}} = 4 \hbar k \tilde{B}  \tilde{C}$. In this case the dynamics could be reproduced instead by coupling the two systems together using a direct interaction Hamiltonian given by $H_{\ms{eff}}$, and then separately coupling the controller to a field via the operator $L$. This kind of \textit{direct coupling} configuration is depicted in Fig.~\ref{fig4}. 

Configuration (b) has the ability to cancel the noise input to both the system and the controller, which is achieved by setting $e^{i(\theta + \phi)} = -1$ and $A = B$ as before, and in addition $M = L$ and $\chi = \theta$. The resulting equations of motion are   
\begin{align} 
    \dot{S} = & i[H_{\ms{s}}/\hbar + 4 k \tilde{B}  \tilde{C}, S]  , \\ 
     \dot{X} =  & i[H_{\ms{a}}/\hbar  + 4 k \tilde{B} \tilde{C}, X ]  . 
     \label{fullH}
\end{align} 
We see that these equations describe two systems coupled together solely by the interaction Hamiltonian $H_{\ms{eff}}$.  Thus from a theoretical point of view the use of field-mediated coupling to connect quantum systems subsumes the use of direct coupling, because the former can simulate the latter. 

The fact that the field coupling configurations can reproduce the direct-coupling scenario in Fig.~\ref{fig4} means that they can extract entropy from the system using a ``state-swapping'' procedure. The effective direct coupling, along with the control of the auxiliary Hamiltonian, can be used to realize a joint unitary operation that swaps the states of the system and controller. This method of control is discussed, for example in~\cite{Wang11, Jacobs15}, and is the mechanism used in resolved-sideband cooling. The latter is presently the state-of-the-art for cooling nano-mechanical oscillators~\cite{Schliesser08, Schliesser09, Teufel11b, Chan11} and the external motion of trapped ions~\cite{Leibfried03, Monz16}. If one prepares the controller in the state in which one wishes to prepare the system, then swapping the state of the system with that of the controller prepares the system in the desired state, automatically transferring any entropy in the system to the controller. In this case, assuming a perfect controller, the fidelity with the which the system can be prepared in the desired state --- that is, the degree of control that can be obtained --- is determined by the speed at which the swap can be implemented. The faster the swap is performed the less time the noise that drives the system has to degrade the state as it is loaded into the system. Of course, to continually re-prepare the system in the desired state the controller must get rid of the entropy it extracts from the system. This is the reason that we include in Fig.~\ref{fig4}(b), in addition to the direct coupling between the system and controller, a coupling between the controller and a field that can act as a zero temperature bath. 

The limit with which a coherent interaction can prepare a system in its ground state in the presence of thermal noise was explored by Wang \textit{et al.\ }in~\cite{Wang13}. There the authors presented a simple analytic expression as a bound on the minimum achievable excited-state probability, for which they provided strong evidence, and which is expected to be valid when the interaction rate is much greater than the thermal relaxation rate. If we denote the interaction Hamiltonian between the system and the controller by $H_{\ms{int}}$, then this bound is $P \geq (\pi/2)(\gamma n_T)/|H_{\ms{int}}|$. For our coherent control configurations, in which the field-mediated coupling is bounded, we have 
\begin{align} 
   |H_{\ms{int}}| = |H_{\ms{eff}}| = 4 k |\tilde{B}\tilde{C}| \leq 8 k . 
\end{align} 
The resulting value of the bound asserted in~\cite{Wang13} is 
\begin{align} 
 P \geq \frac{\pi}{2} \left( \frac{\gamma n_T}{8k}  \right) = \frac{\pi}{2} P_{\ms{MBF}}^{(\infty)} , \;\;\; k \gg \gamma n_T . 	\label{lowb}
\end{align} 
This bound is a factor of $\pi/2 \approx 1.57$ higher than that achievable by MBF. 

We now note that the coherent interaction provided by $H_{\ms{eff}}$ is not the only control mechanism available to the coherent control configurations given in Fig.~\ref{fig3} (a) and (b). By choosing the Hermitian operators $A$ and $B$ appropriately, along with the phases $\theta$ and $\phi$, these configurations can create an effectively dissipative interaction with the field. Further, this can be achieved with the simple feedback loop depicted in Fig.~\ref{fig4} (a) in which we have removed the quantum auxiliary system. The equation of motion for the system in this case, obtained from Eq.(\ref{S1a}) by setting $L=0$, is 
\begin{align} 
    d S = & \, i \! \left[ \frac{H_{\ms{s}}}{\hbar}, S \right] dt - k \mathcal{K}(\tilde{D}) S dt  \nonumber \\ 
     & + \sqrt{2k} ( [S, \tilde{D}^\dagger  ] da_{\ms{in}} - [S, \tilde{D}  ] da_{\ms{in}}^\dagger ) . 
    \label{noc}
\end{align} 
If we now choose $A = \sigma_x$ and $B = \sigma_y$ then 
\begin{align} 
    D = \sigma_x + i\sigma_y = 2 \sigma = 2 |0\rangle \langle 1| . 
\end{align} 
This makes $D$ the decay operator for the qubit and results in damping for the qubit at rate $2k$. If we include thermal noise for the qubit in Eq.(\ref{noc}) then the steady-state population of the excited state, assuming $k \gg \gamma n_T$, is 
\begin{align} 
   P_{\ms{nc}} = \frac{\gamma n_T}{4 k} = 2 P_{\ms{MBF}}^{(\infty)} ,  \;\;\; k \gg \gamma n_T . 
\end{align} 
While this is twice the minimum value for MBF with infinite $\mu$, it doesn't require any quantum auxiliary! 

To summarize the above discussion, we see that the coherent configurations (a) and (b) in Fig.~\ref{fig3} possess two separate mechanisms by which they can prepare the system in its ground state. While the best performance that can be achieved by each of these mechanisms individually is less than that achievable with MBF, presumably both can be used simultaneously, at least to some degree. 

\subsection{Performance of CFC with a single-qubit auxiliary} 
\label{numcfc}

We have obtained some insight into the power of coherent feedback, but we have not found an analytic expression for the maximum performance of CFC under our constraint. To explore this question further we now turn to numerical optimization. For each of the configurations in Fig.~\ref{fig3} the space of possible options for implementing control is large. In (a) for example, we have four Hermitian operators, $A$, $B$, $C$, and $C'$ that we can vary in an essentially arbitrary way with time, as well as the Hamiltonian of the auxiliary system and the phases $\theta$ and $\phi$. The purpose of numerical optimization is to search over the space of functions of time, called the \textit{control functions}, that we can choose for the above quantities in order to maximize the performance of the CFC configuration. To perform such a search we must characterize the control functions in terms of a finite set of parameters. The optimization procedure then performs a search over the space of parameters. A given set of control functions that describes how all of the time-variable quantities change with time, defined over some duration, $T$, is called a \textit{control protocol}. 

A simple way to parametrize our control functions is to divide the time over which the control will be applied, $T$, into $N$ equal intervals, and make the control functions piecewise-constant on these intervals. This is the parametrization we use here. Since we are interested in steady-state control we choose a duration $T$ over which to define the control functions, and we then apply this control repeatedly until the steady-state is obtained. Thus our control protocol will be periodic with period $T$. The total number of real parameters over which we must search is the number of (real-valued) control functions multiplied by the number of intervals $N$. Each Hermitian operator appearing in the equation of motion for an $M$ dimensional system is defined by $M^2 - 1$ real parameters (one less than $M^2$ because the motion generated by an operator is unaffected by adding to it a multiple of the identity). 

For our numerical exploration of the performance of CFC we use only the simplest auxiliary system for the controller, namely a single qubit. With this choice, and allowing the energy gap of the system as an additional control function, the total number of parameters for configurations (a), (b), and (c), in Fig.~\ref{fig3} are $18N$, $25N$, and $24N$, respectively.  For configurations (a) and (b) in Fig.~\ref{fig4} there are $7N$ and $15N$ parameters, respectively. Note that in Fig.~\ref{fig4} (b) the operators $A$ and $C$ only appear in the evolution as a product, so that together they require 5 rather than 6 parameters. Further details of the procedure we use for numerical optimization are given in the Appendix. 

For the numerical analysis we must choose values for the parameters for our control problem, and we use the same ones we used for the simulation of MBF in Section~\ref{secmbf}, namely $\gamma = 0.01k$ and $n_T = 1$. Recall that the coherent feedback control problem is completely specified by these two parameters when we measure $\gamma$ in units of $k$. We first perform a numerical search for protocols that employ the direct-coupling configuration (Fig.~\ref{fig4}(b)). For the chosen values of $\gamma$ and $k$ the lower bound for direct coupling, given in Eq.(\ref{lowb}), is $P = 0.00196$. The best we are able to achieve for the direct coupling configuration using numerical optimization is $P = 0.0031$. It is expected that this should be higher than the lower bound in Eq.(\ref{lowb}) as this bound is not expected to be achievable in the steady-state but only for preparation at a single instant. Our results are thus consistent with the claims of~\cite{Wang13}. 
 
Turning to the field-mediated CFC protocols in Fig.~\ref{fig3} we first evaluate the performance of (a) and (b). The results of our numerical optimization, presented in detail in the Appendix, indicate that the best possible performance of both (a) and (b) is exactly that same as that of the continuous MBF protocol with infinite feedback force, which for the given parameter values is $P = 0.00125$. This result is interesting for at least two reasons. First, it seems somewhat remarkable that the CFC protocol is able to perform as well as the MBF protocol when its feedback interaction forces are limited precisely to those of its measurement interaction forces, while the MBF protocol can use infinitely fast Hamiltonian control. Second, it would seem rather coincidental that a two-level auxiliary would exactly match the performance of MBF unless this performance is a bound that is independent of the size of the auxiliary. This suggests that the bound previously established for MBF with infinite feedback, Eq.(\ref{infmu}), may be the upper limit for any control under a bound on the Markovian coupling to a field. Certainly this would appear to be an interesting question for future work. The fact that (a) and (b) given the same performance suggests that the addition of the interaction given by $M$ in (b) may not provide any additional power for the control process, at least when the dynamics of the auxiliary is unconstrained. This too may be a question worth pursuing in future work. 

Finally we evaluate the performance of configuration (c) in Fig.~\ref{fig3}. We find that this configuration is not able to provide \textit{any} control over the entropy of the qubit, in that it is unable to reduce the probability of the excited state below the uncontrolled value of $P = 1/3$. This shows that the ability to cancel the input noise to the system, a property possessed by configurations (a) and (b), is essential for performing non-trivial control when the interactions with the system and the fields are Hermitian (and thus non-dissipative), and when the auxiliary has only two levels. It seems likely that this will remain true for couplings that are dissipative but in which the field is at the same temperature as the noise one wishes to control. We expect however, that configuration 1(c) will be able to perform non-trivial control when the auxiliary has more than two levels, since such a scenario should be able to mimic the functioning of MBF. 

\begin{figure}[t] 
\leavevmode\includegraphics[width=1\hsize]{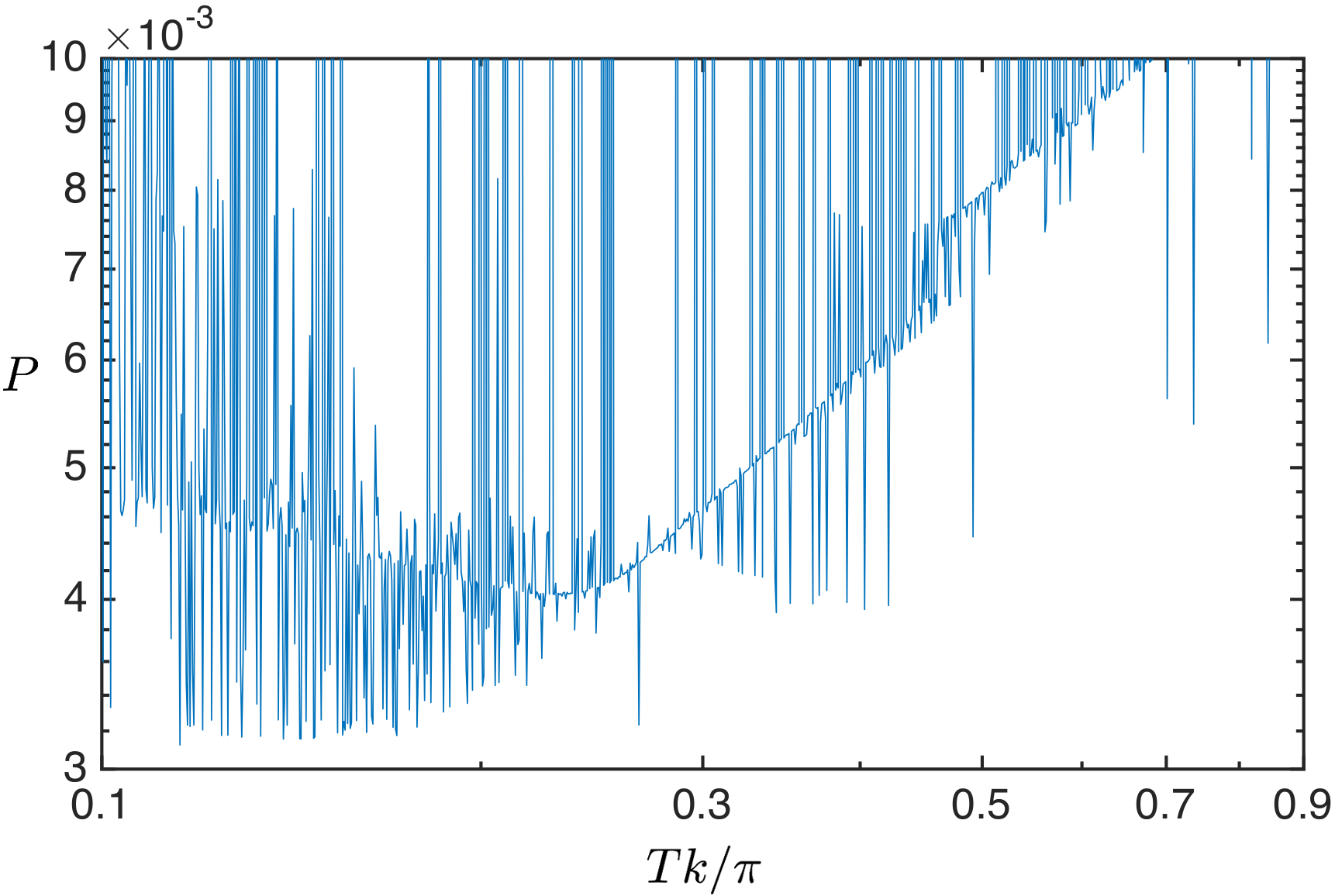} 
\caption{Numerical optimization for the direct-coupling configuration of Fig.~\ref{fig4}(b). Here we plot the result of 1000 independent searches for a range of values of the protocol period, $T$, logarithmically equally spaced between $T = \pi/(10k)$ to $T = \pi/k$. (The plot actually shows the subset of  results that lie within the slightly smaller interval $T k/\pi \in [0.1,0.9]$.) We allowed the control functions to have $N=8$ piecewise-constant segments with the duration $T$. Similar results were obtained using $N=4$. The minimum value  obtained is $P = 0.0031$. } 
\label{fig5} 
\end{figure} 

\begin{figure}[t] 
\leavevmode\includegraphics[width=1\hsize]{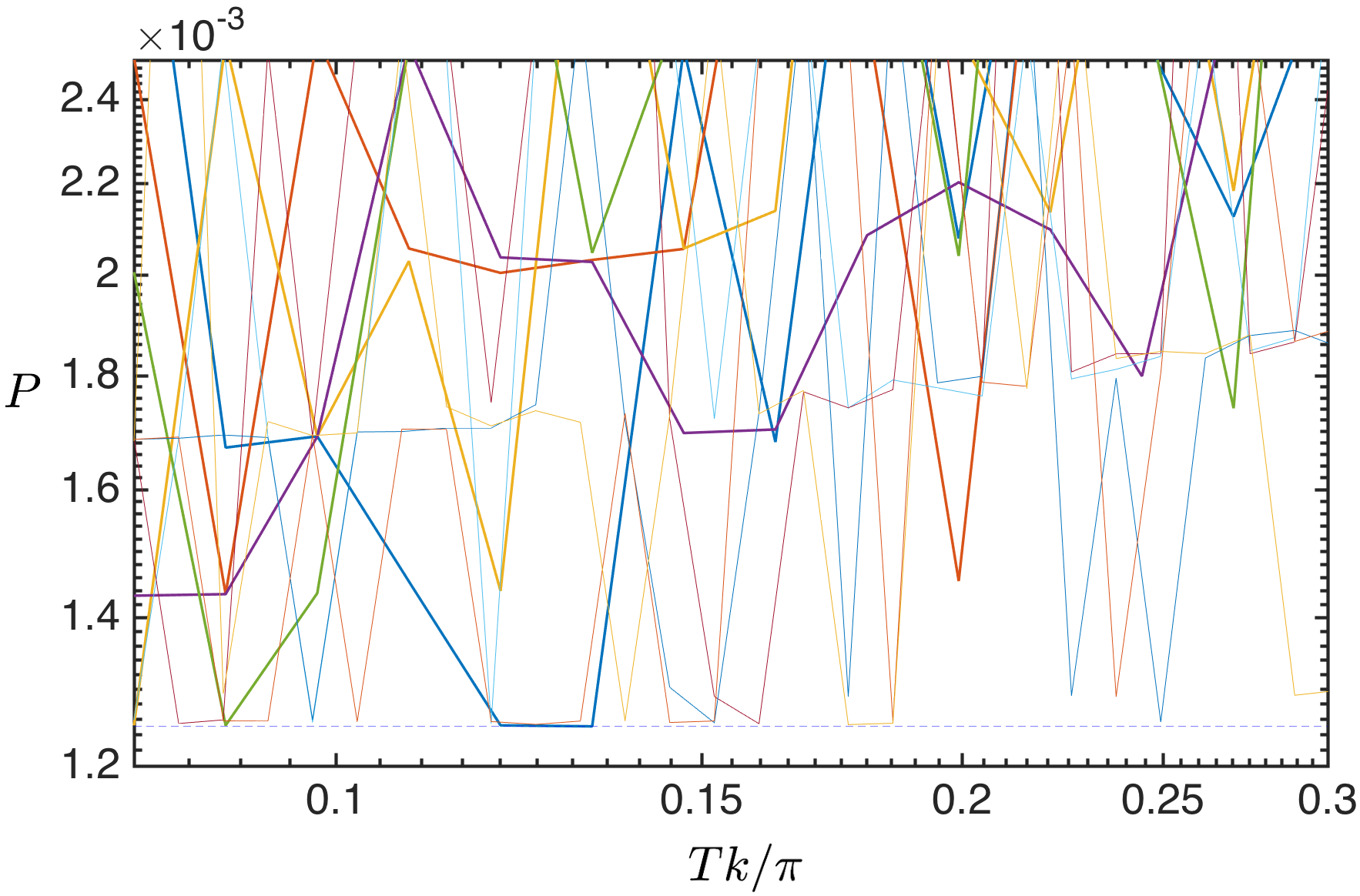} 
\caption{Here we show the results of a numerical search for the field-coupling control configurations depicted in Fig.~\ref{fig3}(a) and (b). The thick lines show the results for (a) and the thin lines the results for (b). In both cases we see that we find a number of protocols for which $P$ is very close to the minimal MBF value of $0.00125$, indicated by the dashed line, but find no protocols that achieve a lower value.} 
\label{fig6} 
\end{figure}

\section{Conclusions}  
\label{conc}

We have identified two distinct, and rather general, configurations for implementing field-mediated coherent feedback control (those depicted in Fig.~\ref{fig3}(b) and (c)), and used numerical search methods to compare their performance in controlling a single qubit to that of continuous measurement-based control when i) the interactions with the field are subject to the same constraints for both forms of control, and ii) the auxiliary for the CFC is a two-level system, and iii) the interactions with the system are via Hermitian operators, and are thus non-dissipative and correspond to the measurement of observables. 

While numerical search methods cannot guarantee to find the best possible performance, our results indicate that i) CFC configuration 1(b) is essentially useless for control when the auxiliary has only two levels; ii) CFC configuration 1(b), and its simpler version 1(a), are able to exactly match the performance of MBF when the latter has access to infinite feedback forces. This suggests that this performance may be fundamental limit under a constraint on the size of the coupling to a field, although certainly further investigations will be required to determine this question one way or the other. 
 
We have also considered CFC implemented by a direct coupling between the system and auxiliary, where the size of the direct coupling is restricted to that which can be obtained as an effective coupling by employing the field-mediated CFC configurations. The bound on this direct coupling scenario asserted previously in~\cite{Wang13} indicates that it cannot achieve the performance that we obtained for the field-mediated protocols, and the numerical results we obtained provide further conformation of this assertion. 

Our analysis here does not constitute a comparison of the performance of CFC and MBF as far as fully \textit{practical} considerations are concerned, since we have assumed ideal controllers in both cases. In reality the performance of MBF will be limited by the efficiency of the measurement and any time-delay and bandwidth constraints of the processing. Similarly the performance of CFC will be limited by the noise to which the auxiliary quantum system is subject to, be this thermal noise, damping, or dephasing. Which method of control is more powerful may well depend on the physical setting of an implementation. Our results have raised a number of further questions about the relative power of various control scenarios, and the ultimate limits to control that employs field-mediated coupling. In addition to these questions that of the relative performance of CFC and MBF for non-ideal controllers is also a line of investigation that we think is worth pursuing. 

%*** big difference between a,b and c. The latter must be able to provide control with a large enough auxiliary because the classical controller has no coherence either. It seems likely then the inability to effect control is due to the small size of the auxiliary. 
%
%*** CFC with no controller is able to do well 
%
%*** CFC is able to match MBF with only a small auxiliary and a bounded interaction strength. 
%
%*** CFC does not do any better than MBF with infinite mu, indicating that this may be the best possible performance for bounded interaction rate. 
%
%*** definitely more questions to ask
%
%*** we have assumed perfect controllers, both classical and quantum. It would also be interesting to compare the performance of MBF and CFC when the controllers are subject to imperfections. 
%
%*** we have also only used the simplest auxiliary systems 

\appendix 

\vspace{-3mm}
\section{Numerical optimization for coherent feedback}
\vspace{-2mm}

To simulate the CFC configurations we must use the master equation rather than the quantum noise equations that we presented in the main text. The master equations look a little different than the quantum noise equations because the latter have been specialized so that they apply either to a system operator or an auxiliary operator. This simplifies the form of the quantum noise equations for both cases because system operators, when evaluated at a given time, always commute with auxiliary operators when evaluated at the same time.  The master equation however must be able to evolve simultaneously a joint state of both system and auxiliary, which is equivalent to being able to evolve both system and auxiliary operators. The master equation for the configuration in Fig~\ref{fig3}(a) is~\cite{Jacobs14} 
\begin{align}
   \dot{\rho} = &\, \frac{-i}{\hbar} [H_{\ms{s}} + H_{\ms{a}}, \rho] + 2[L^\dagger , Ae^{i\theta} \rho]  - 2 [L , \rho A e^{-i\theta}]  \nonumber \\  
   & +  2 [B ,   L e^{i\phi} \rho - \rho L^\dagger e^{-i\phi}]  - (\mathcal{K}[D] + \mathcal{K}[L]) \rho \nonumber \\ 
   & - \frac{\gamma}{2} \left[ (n_T + 1) \mathcal{K}(\sigma) + n_T \mathcal{K}(\sigma^\dagger) \right] \rho , 
\end{align} 
in which $\rho$ is the joint density matrix for the system and controller and the other operators and superoperators are defined in the main text. The master equation for configuration 1(b) is 
\begin{align}
      \dot{\rho} = &\, \frac{-i}{\hbar} [H_{\ms{s}} + H_{\ms{a}}, \rho] + 2[V^\dagger , Ae^{i\theta} \rho]  -  2[V , \rho A e^{-i\theta}]  \nonumber \\  
      & +  2[M^\dagger , Be^{i\chi} \rho]  -  2[M , \rho B^\dagger e^{-i\chi}]   \nonumber \\ 
   & +  2 [B ,   L e^{i\phi} \rho - \rho L^\dagger e^{-i\phi}] - (\mathcal{K}[D] + \mathcal{K}[V]) \rho \nonumber \\ 
   & - \frac{\gamma}{2} \left[ (n_T + 1) \mathcal{K}(\sigma) + n_T \mathcal{K}(\sigma^\dagger) \right] \rho , 
\end{align} 
and that for 1(c) is 
\begin{align}
      \dot{\rho} = &\, \frac{-i}{\hbar} [H_{\ms{s}} + H_{\ms{a}}, \rho] - (\mathcal{K}[A] + \mathcal{K}[B] + \mathcal{K}[L] + \mathcal{K}[M]) \rho \nonumber \\  
      & + 2[M^\dagger , Ae^{i\theta} \rho]  - 2 [M, \rho A e^{-i\theta}]   \nonumber \\ 
      & + 2 [B , Le^{i\phi} \rho - \rho L^\dagger e^{-i\phi})]   \nonumber \\
   & - \frac{\gamma}{2} \left[ (n_T + 1) \mathcal{K}(\sigma) + n_T \mathcal{K}(\sigma^\dagger) \right] \rho , 
\end{align} 
As discussed in the main text, for each of these configurations we allow each operator, with the exception of the system Hamiltonian $H_{\ms{s}}$ to be a piecewise-constant function on a fixed time interval $T$. Since we want to achieve steady-state control we apply this protocol of duration $T$ repeatedly to obtain a periodic control protocol. We simulate this control protocol for long enough to determine the steady-state. So long as we allow enough piecewise-constant segments within the duration $T$ the precise value of $T$ should not matter. 

A basic numerical search involves choosing the period $T$ and number of segments $N$ and, beginning at a random point in the search space, searching over the space of piecewise-constant functions that we can choose for all the operators and phases appearing in the master equation so as to minimize the steady-state value of the excited-state probability, $P$. We employ this basic search approach in two ways. The first is to perform an independent search for each of a range of values of $T$. The second is inspired by the ``path-tracing'' method which was introduced by Moore-Tibbetts~\textit{et al.\ }in~\cite{Moore12b} and revealed as a powerful tool for time-optimal control in~\cite{Jacobs16b}. In this approach we begin by choosing a random point in the search space and performing a search for a given value of $T$. We then reduce the value of $T$ by some predefined amount and perform the search again, but this time starting from the point in the search space --- that is, the control protocol --- found by the previous search at the somewhat larger value of $T$. We then repeat this procedure each time reducing the value of $T$. Both the first and second approach provide a series of control protocols for a range of values of $T$, but the second has been shown to be significantly more effective, for at least some problems, in finding protocols in which the duration $T$ is an important constraint. The intuition here is that the shorter the time required to transfer the entropy from the system to the controller, the better will be the final steady-state, since the thermal noise introduces entropy into the system at a fixed rate. 

For the direct-coupling protocol depicted in Fig.~\ref{fig4}(b) we find that the first method is just as effective as the second, and we plot in Fig.~\ref{fig5} the results of a single scan of 1000 points covering the interval $T = [0.1,1]\pi/k$. We see from this that while not every search finds a good protocol (that is, a near-optimal value of $P$), a significant fraction of them do.  

For the field-mediated configurations (Fig.~\ref{fig3}(a) and (b)) we find that the second method is superior. For (a) we performed 5 scans starting with values for T in the range $T \in [0.5, 0.55]\pi/k$ and stepping down to $T = 0.08 \pi/k$ using 20 logarithmically equally spaced points. For these runs we used $N=8$ segments within the duration $T$. Twenty scans with $N=4$ segments produced similar results. We followed a very similar procedure for configuration (b), using the same scanning window, this time with $N=4$ and performing 5 scans using $40$ points for each scan. The results for both (a) and (b) are shown in Fig.~\ref{fig6}. 

%Details for Fig.~\ref{fig6}:  (a) 5 trace runs, 20 steps per run, starting T values is range 0.5 pi to 0.55 pi, ending T value is 0.08 pi. time segments per protocol is 8, protocol is repeated 40 times to ensure steady-state performance. Also did 20 runs with 4 segments per protocol and obtained very similar results. 

%\bibliography{report_QMT}


\end{document}